\documentclass{PoS}
\let\ifpdf\relax

\usepackage{graphicx,times,amsmath,natbib,epsfig}
\usepackage{aas_macros} 
\usepackage{grffile}

\def\apj{{ApJ}}
\def\apjs{{ApJS}} 
\def\apjl{{ApJL}}
\def\aap{{\em A.\&A}}

\def\mnras{{MNRAS}}

\def\prd{{Phys Rev D}}

\title{Epoch of Reionization modelling and simulations for SKA}

\ShortTitle{EoR modelling for SKA}

\author{\speaker{Ilian T. Iliev}\\ 
        Astronomy Centre, Department of Physics \& Astronomy, Pevensey II 
Building, University of Sussex, Falmer, Brighton BN1 9QH, United Kingdom\\
        E-mail: \email{I.T.Iliev@sussex.ac.uk}}

\author{Mario G. Santos\\
	Department of Physics, University of Western Cape, Cape Town 7535, South Africa\\
	SKA SA, 4rd Floor, The Park, Park Road, Pinelands, 7405, South Africa\\
	CENTRA, Instituto Superior T\'{e}cnico, Universidade de Lisboa, Lisboa 1049-001, Portugal\\
	E-mail: \email{mgrsantos@uwc.ac.za}}

\author{Andrei Mesinger \\
         Scuola Normale Superiore, Pisa, Italy\\
        E-mail: \email{andrei.mesinger@sns.it}}

\author{Suman Majumdar, Garrelt Mellema\\
Department of Astronomy \& Oskar Klein Centre, AlbaNova,
  Stockholm University, SE-106 91 Stockholm, Sweden\\
 E-mail: \email{smaju@astro.su.se,garrelt@astro.su.se}}

\abstract{In this chapter we provide an overview of the current status of the
simulations and modelling of the Cosmic Dawn and Epoch of Reionization.
We discuss the modelling requirements as dictated by the characteristic 
scales of the problem and the SKA instrumental properties and the planned
survey parameters. Current simulations include most of the relevant physical 
processes. They can follow the full nonlinear dynamics and are now reaching 
the required scale and 
dynamic range, although small-scale physics still needs to be included at 
sub-grid level. However, despite a significant progress in developing novel 
numerical methods for efficient utilization of current hardware they remain 
quite computationally expensive. In response, a number of alternative 
approaches, particularly semi-analytical/semi-numerical methods, have been 
developed. While necessarily more approximate, if appropriately constructed 
and calibrated on simulations they could be used to quickly explore the vast 
parameter space available. Further work is still required on including some 
physical processes in both simulations and semi-analytical modelling. This 
hybrid approach of fast, approximate modelling calibrated on numerical 
simulations can then be used to construct large libraries of reionization 
models for reliable interpretation of the observational data.}

\FullConference{
Advancing Astrophysics with the Square Kilometre Array\\
June 8-13, 2014\\
Giardini Naxos, Italy}

\begin{document}

\section{Introduction}

The Cosmic Dark Ages, Cosmic Dawn and Epoch of Reionization, span altogether 
about a billion years between the CMB last scattering surface at redshift 
$z\sim1100$ and the complete ionisation of hydrogen in the inter-galactic 
medium (IGM) at $z\sim6$. This phase is quite distant from us and difficult 
to study, remaining one of the last poorly understood epochs of the Universe. 
During this time, the very first stars and galaxies formed, then gradually 
ionized the intergalactic medium and enriched it with metals, thereby laying 
the foundations for the cosmic structures we see today. Better knowledge of 
these epochs is therefore of key importance for understanding galaxy formation 
and evolution.

The main obstacle to further progress is the scarcity of observational data, 
which currently mostly probes the tail-end of reionization (both Ly-$\alpha$ 
source surveys and IGM absorption lines probe low neutral fractions, due
to the high optical depth to resonant scattering of such radiation by even 
small amounts of neutral hydrogen) or are integral measures of its history 
(Cosmic Microwave Background, CMB, optical depth and polarization; kinetic 
Sunyaev-Zel'dovich effect, kSZ; Near Infrared Background, NIRB). The redshifted 
21-cm signal promises to provide full 3D tomographic observations of the 
intergalactic medium throughout, and possibly even before reionization. The 
first generation of experiments, currently ongoing, will most likely provide 
just a statistical detection of the signal, e.g. power spectra, variance 
and/or probability distribution functions (PDFs) and their higher moments. 
In contrast, SKA should be able to also perform imaging and 3D mapping, as well 
as go much deeper in redshift, due to its far superior sensitivity, thereby
completely transforming reionization research. 

The shortage of observational constraints has meant that simulations have 
played and continue to play a larger role than in other areas, since we need 
to rely on modelling to understand even the basic properties of the expected
signals. This in turn steers the design of the observational experiments 
targeted at this science. Furthermore, better understanding of the expected 
21-cm signatures and their cross-correlations with other observations is 
important for reliably confirming any detection, since the signal is weak 
compared to the foregrounds which are several orders of magnitude stronger. 
Separating the signal and the foregrounds will not be trivial (see e.g. 
Chapman et al. 2015, reference: PoS(AASKA14)005) and subtraction may leave 
contaminating residuals mimicking the signal. The 
simulations will have a further important role to play in the interpretation 
of any detections. The reason for this is the disconnect in scales between 
the objects we want to study, i.e. the first galaxies, which are very small 
and faint, thus largely impossible to observe directly, and the actual 21-cm 
and other signals we will detect, which typically relate to large-scale 
patterns such as the reionization patchiness. The latter are caused by large 
number of clustered, individually dim sources. Therefore, the interpretation 
of any signals in terms of the nature and properties of the first galaxies is 
non-trivial and requires detailed modelling.
 
In recent years two basic approaches emerged in modelling the reionization 
patchiness - either by direct numerical simulations 
\citep[e.g.][]{1997ApJ...486..581G,2000MNRAS.314..611C,2000ApJ...535..530G,
2002ApJ...575...33R,2006MNRAS.369.1625I,2014MNRAS.439..725I} 
or through semi-analytical/semi-numerical modelling 
\citep[e.g.][]{2004ApJ...613....1F,2007ApJ...654...12Z,2009ApJ...703L.167A,
MF07,CHR09,2010MNRAS.406.2421S,2013ApJ...776...81B}, 
along with some intermediate methods involving simplified simulations 
\citep{2007ApJ...657...15K,2009MNRAS.393...32T}. These different lines of 
attack are in practice highly complimentary and each has its advantages as 
well as disadvantages, as discussed in more detail in the next section.

\section{Overview of current modelling}

\subsection{Radiative transfer simulations}

\begin{figure}
\includegraphics[width=.5\textwidth]{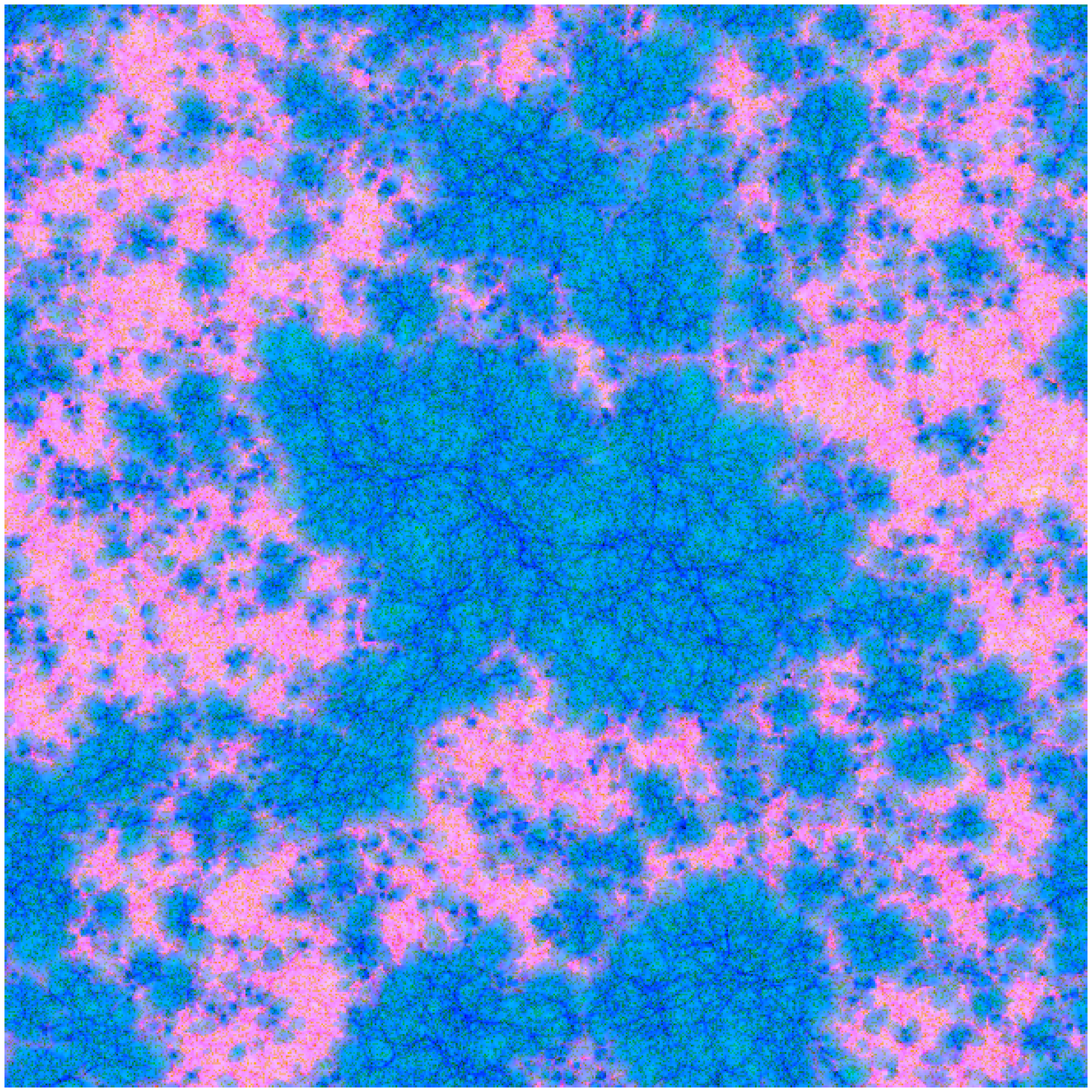}
\includegraphics[width=.5\textwidth]{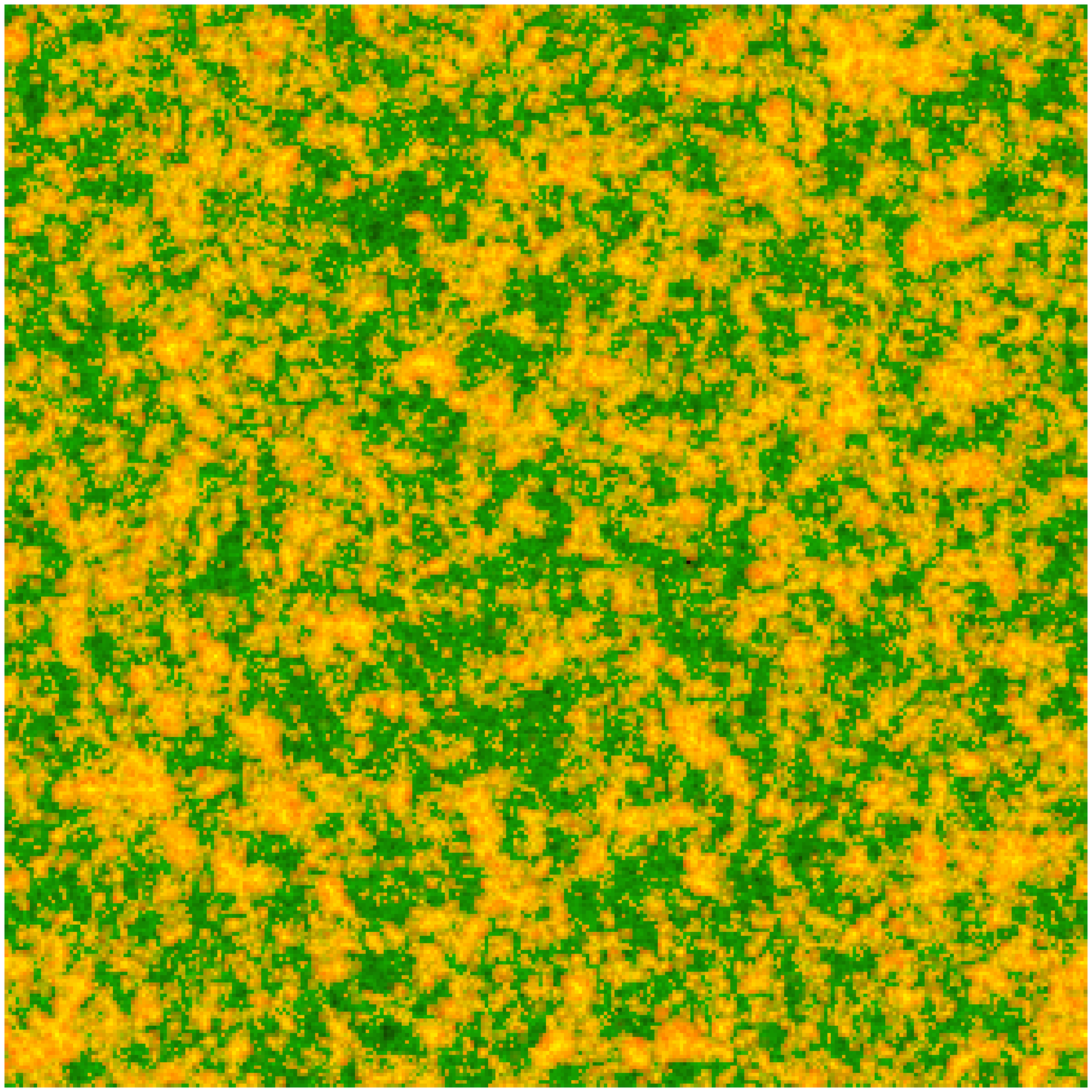}
\caption{Spatial slices from radiative transfer simulations of reionization.
Shown are the ionized (orange/blue) and neutral gas density (green/pink) 
at box-averaged ionized fractions by mass of $x_m\sim0.4$ for (left) a small, 
high resolution volume ($47\,Mpc/h$, $612^3$ radiative transfer grid) and a 
very large volume, comparable to the expected full FOV for SKA1-LOW 
($500\,Mpc/h$, corresponding to $\sim5^\circ\times5^\circ$ on the sky, $250^3$ 
grid) (Dixon et al., in prep.).}
\label{images:fig}
\end{figure}
\begin{figure}
\includegraphics[width=\textwidth]{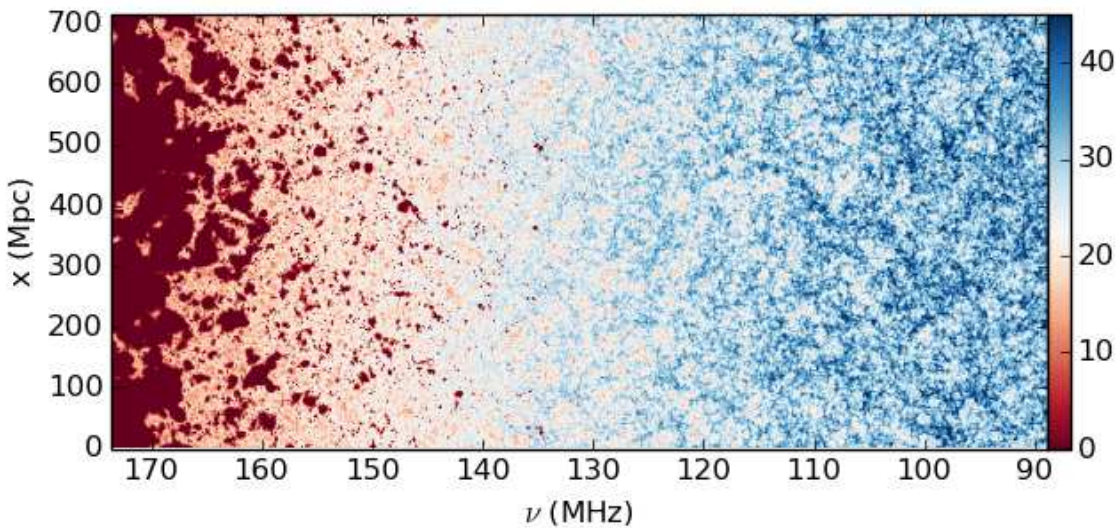}
\caption{Space-frequency slice of the 21-cm emission from the intergalactic
medium showing the evolution of the reionization geometry over time, from 
fully-neutral (blue) to mostly ionized (green). This is based in a radiative 
transfer simulation with $500\,\rm Mpc/h$ volume and $300^3$ cells. The color 
shows the differential brightness temperature in mK at the full simulation 
resolution. (Dixon et al., in prep.)}
\label{evolution:fig}
\end{figure}

The simulations of cosmic reionization fall in two broad types, as a
consequence of the huge range of scales involved, which in practice 
cannot be covered in a single simulation on current computer hardware. 
The first type consists of small volume, high-resolution simulations which 
can be used to study in detail the formation of early galaxies and their 
radiative and supernova feedback on the gas with fully-coupled radiative 
hydrodynamic simulations. Such simulations cannot capture a large enough 
volume to represent the global reionization process or its observational 
signatures \citep[e.g.][]{1997ApJ...486..581G,2000MNRAS.314..611C,
2000ApJ...535..530G,2002ApJ...575...33R}. The second type consists of 
large-scale simulations, which instead follow volumes sufficiently large to 
study the global evolution, but lack resolution to directly resolve the 
small-scale physics \citep[e.g.][]{2006MNRAS.369.1625I,2007ApJ...671....1T,
2014MNRAS.439..725I} (see Figs.~\ref{images:fig} and \ref{evolution:fig}). 
This type of simulations decouple the radiative transfer, star formation and 
feedback processes from the underlying structure formation. The latter is done 
first, typically using large-scale N-body simulations, which is then followed 
up by detailed radiative transfer and non-equilibrium chemistry simulations.
This approach allows for much larger dynamic range, and thus the larger 
volumes required for understanding the global reionization process and its 
observational signatures. In this case the detailed physical processes at 
unresolved scales have to be included using subgrid recipes, which are 
themselves based on high-resolution simulations or other modelling 
\citep[e.g.][]{2007MNRAS.376..534I,2007MNRAS.377.1043M,2012ApJ...756L..16A,
2014arXiv1407.2637A}. Both types of simulations have played a major role in 
the significant recent advances in this area. We now have a good understanding 
of the characteristic scales of the process \citep{2011MNRAS.413.1353F,
2014MNRAS.439..725I}, the physical processes affecting the patchiness and the 
various observational features which might help to discriminate between 
different EoR models \citep[e.g.][]{2008ApJ...685...40W,2009A&A...495..389B,
2012ApJ...756L..16A,iliev12,2012ApJ...747..126A,2012ApJ...760....4O,
2013ApJ...769...93P,Jensen2013,sobacchi14}.

\subsection{Semi-numerical simulations}   
\label{seminanalyt:sect}

Some of the early approaches to characterise the 21cm signal from the epoch 
of reionization have relied on more simple analytical models, mostly based
on the excursion-set approach. These can be very useful to quickly generate 
the signal, in particular the power spectrum of 21-cm brightness temperature 
fluctuations \citep{2004ApJ...613....1F,2006MNRAS.365..115F,
2005MNRAS.363..818S}. The speed and ease of use, allows for straightforward 
tests
of the ability of a given experiment to constraint cosmological 
and astrophysical parameters \citep{2006PhRvD..74h3517S,2006ApJ...653..815M,
2008PhRvD..78b3529M}
and more generally to rapidly explore the vast parameter space available. The 
analytical models have also been useful to understand the possible 
contributions to the 21 cm signal at high redshifts \citep{2005ApJ...624L..65B,
2007PhDT.......197P}
, although it is at low redshifts ($z\lesssim 10$) that they seem to provide 
a better description of the 21-cm 2-point correlation function 
\citep{2008ApJ...689....1S}.
However, these models have several issues in properly dealing with the spatial distribution 
of the reionization process, such as bubble overlap and ignore complicated 
astrophysics during reionization. Such complexities are better handled in full 
(but expensive) numerical simulations, as was discussed above.

\begin{figure}
\includegraphics[width=\textwidth]{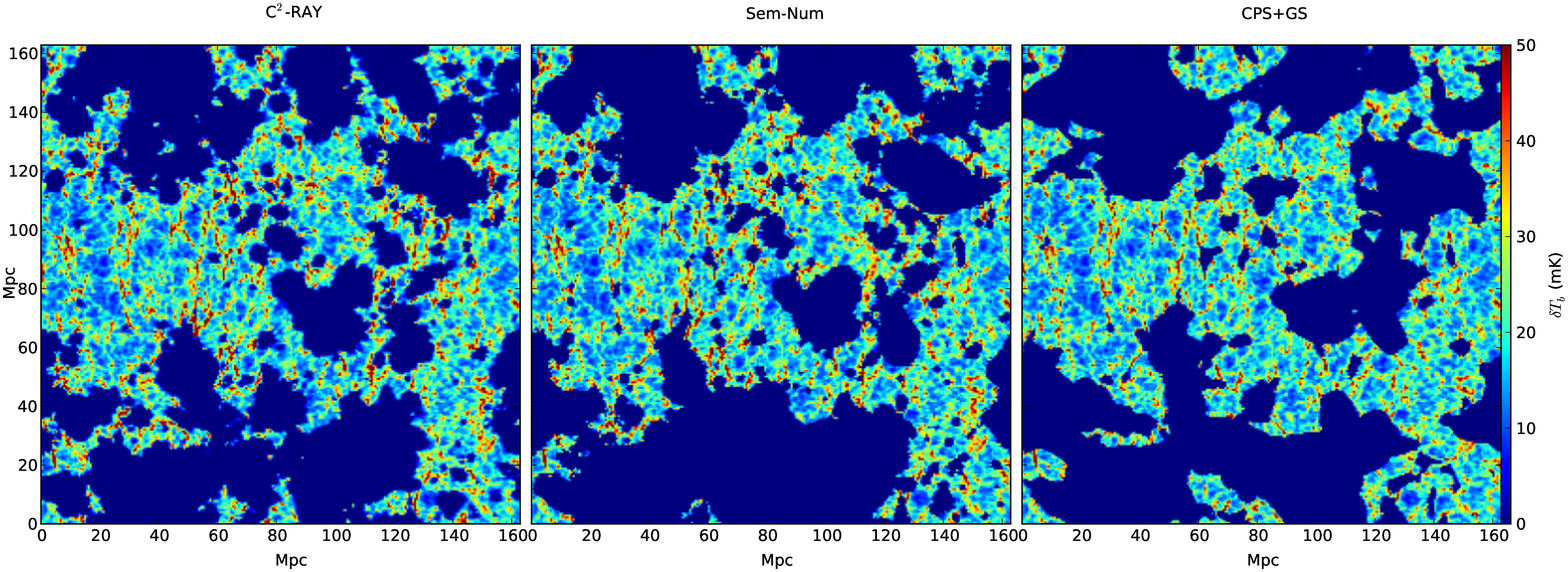}
\includegraphics[width=\textwidth]{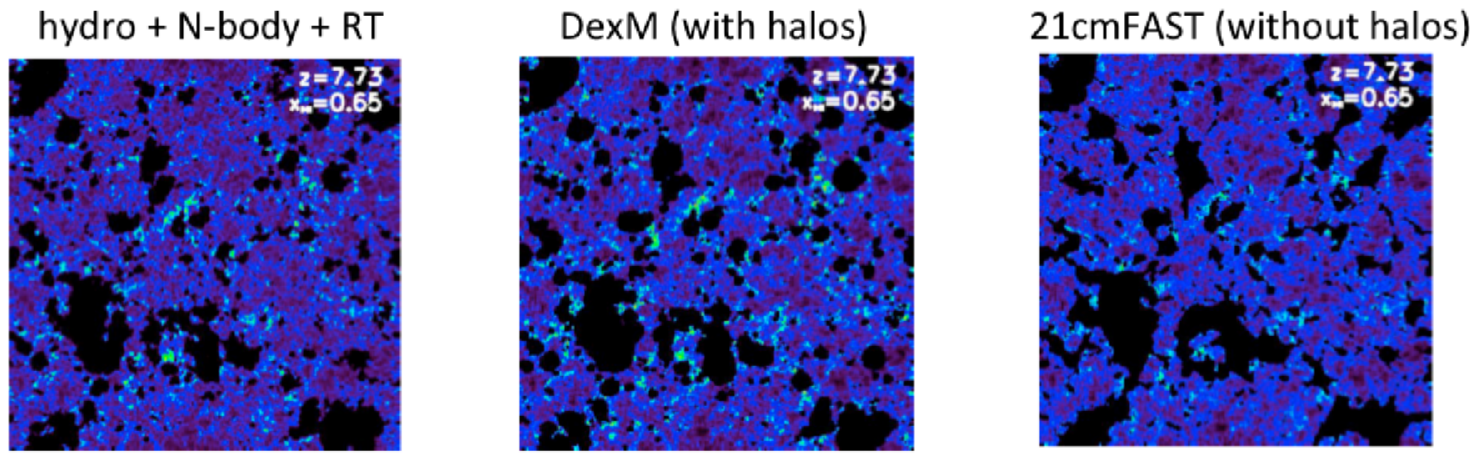}
\caption{
  Reionization morphology through brightness temperature maps in
  redshift space for radiative transfer simulation (left) vs. two
  semi-numerical approaches - one which takes the halo positions into
  account (middle) and one that does not (right). For details see
  the original papers \citet{majumdar14} (top) and \citet{2011MNRAS.411..955M} 
  (bottom).  }
\label{seminum_vs_sim:fig}
\end{figure}

An intermediate approach that has become more popular are the so called 
semi-numerical 21cm simulations which try to merge the speed and ease of use 
of analytical models with the ability to follow the spatial evolution of 
reionization in more detail as provided by numerical simulations. The basic 
feature of these semi-numerical simulations is that they start from a random
Gaussian field realization of the cosmological density field, or a full N-body
simulation data, but then replace the time-consuming radiative transfer (RT) 
with an excursion-set approach \citep{2004ApJ...613....1F} that determines if 
a region is ionized by comparing the number of ionizing photons to the number 
of atoms (plus recombinations) inside it. As in conventional cosmological RT 
simulations, this algorithm can be applied to discrete halo source 
fields, obtained either from traditional N-body simulations 
\citep[e.g.][]{2007ApJ...654...12Z,
CHR09,2011MNRAS.414..727Z}
or through faster methods involving excursion-set formalism and perturbation 
theory \citep{MF07,2010MNRAS.406.2421S,2011MNRAS.411..955M,2009ApJ...703L.167A,
GW08}.
The latter approach results in some reduction in accuracy 
\citep[see Fig.~\ref{seminum_vs_sim:fig}; ][]{2011MNRAS.414..727Z,majumdar14}, 
but substantially increases the achievable dynamic range. Some of these 
approaches have also been extended to simulating the earlier epochs of the 
21cm signal: the 
Cosmic Dawn/Dark Ages \citep{2011MNRAS.411..955M,2010MNRAS.406.2421S}.
During the Dark Ages, the 21cm signal is governed by soft UV and X-ray photons, 
which can have mean free paths of hundreds of Mpc, requiring large volumes, 
which makes it a challenge to implement this in a complete radiative 
transfer/gas dynamics algorithm.

Semi-numerical simulations of reionization have been extensively tested 
against numerical simulations, with good agreement on moderate to large 
scales ($>$Mpc), relevant for most SKA 
science \citep{2011MNRAS.411..955M,2011MNRAS.414..727Z,majumdar14}.
The approximations start to break down however on non-linear scales, and care 
should be taken by comparing against more accurate methods when extending their 
application to any non-standard problems.  Nevertheless, their speed and 
efficiency allows for rapid exploration of the large-parameter space of 
uncertainties, ushering in an exciting era of 21cm astrophysical parameter 
studies \cite[e.g.][]{ME-WH14,Pober13}.

The analysis of \citet{2011MNRAS.414..727Z} has been largely focused on 
the comparison of the spherically average power spectrum of the 21-cm 
signal from different simulations. Their study suggested that
these semi-numerical techniques can mimic the 21-cm power spectrum
from that of a radiative transfer simulation with $\geq 70\%$ accuracy
at length scales relevant for the present and future surveys. This
study was extended in more recent works \citep{2011MNRAS.411..955M,majumdar14} 
to include the comparison of EoR history, 
the redshift space anisotropies in the power spectrum and the morphology 
of the ionization fields (Fig.~\ref{seminum_vs_sim:fig}). Those analyses 
suggest that the halo based semi-numerical simulations can predict the 
history of EoR with an accuracy of $\ge 90\%$ when compared with radiative 
transfer simulations for length scales $k \leq 1.0\,{\rm Mpc}^{-1}$, whereas 
the same accuracy for the Press-Schechter based technique is $\ge 40\%$ at 
same length scale range. They have also shown that the anisotropies in the 
power spectrum of the 21-cm signal due to redshift space distortions can also 
be mimicked by both type of the semi-numerical simulations with an accuracy 
of 85\% or more, which is well within the noise uncertainties of $\sim2000$~hr 
LOFAR or $\sim1000$~hr SKA1-LOW observation at 150 MHz, provided that the actual 
peculiar velocities are used to introduce the redshift space distortions rather 
than using a perturbative technique. It is also found that the cross-correlation 
between the ionization fields from the halo based semi-numerical technique and 
the radiative transfer simulations is more than $85\%$ during almost the entire
period of EoR, for length scales $k \leq 1.0\,{\rm Mpc}^{-1}$, whereas
the same correlation between the conditional Press-Schechter formalism
and the radiative transfer simulations, while good in some regimes, can 
reach values as low as $\sim 10\%$, specifically at late stages of EoR for 
the same length scale range.

These comparisons however, are also limited as the reionization
scenario/model that has been compared in these cases is relatively
simplistic in nature. The reionization history and the nature of the
21-cm signal from this epoch may differ depending on various other
factors. One such important factor is the effect of radiative feedback
on the star formation in low mass halos ($\leq 10^9\, {\rm M_{\odot}}$
in mass). This will most likely affect their star formation
efficiency, although the details remain unclear \citep{couchman86,
rees86, efstathiou92, thoul95, thoul96, gnedin00b,kitayama00,
dijkstra04, hoeft06, okamoto08}. There has been some effort to include
this in numerical \citep{2007MNRAS.376..534I,iliev12,2014arXiv1407.2637A} 
as well as in semi-numerical simulations \citep{sobacchi13}. 
Similarly, another such important issue is the effect of enhanced
recombination in the Lyman Limit Systems and other small-scale structures, 
which are not resolvable in any of these simulations and thus require 
detailed sub-grid modelling (\citealt{MH_sim,CHR09,sobacchi14}, 
Koda et al. in prep., Shukla et al., in prep.). Also, the effect of X-ray 
heating on the spin temperature evolution at the very early stages of 
reionization \citep{mesinger13,2014arXiv1406.4157G} is another important 
factor which has not been included in these comparison studies.

\section{SKA simulations}

\subsection{Basic Simulation Requirements}

The volume and resolution required for proper numerical modelling of the 
reionization process are dictated by both its intrinsic characteristic 
scales (especially the typical sizes of ionized and neutral patches) and 
the instrumental parameters (FOV, beam size and bandwidth). The simulation
volume should be large enough to faithfully sample all relevant scales, 
while the numerical resolution should be such that every beam/bandwidth 
is sampled with a sufficient number of elements so as to avoid discretization 
effects and other numerical artifacts.


The nominal SKA1-LOW Epoch of Reionization survey at the frequencies relevant 
for reionization is proposed to have FOV (FWHM) of $\sim3$ (at $\nu=200$~MHz) 
to 10 degrees (at $\nu=50$~MHz), which corresponds to $\sim500$~Mpc-1.5~Gpc, 
with maximum resolution of $\sim1$~arcmin, or roughly $0.5$~Mpc. Therefore 
simulations of the full SKA1-LOW FOV with sufficient resolution should have 
volume of at least 500~Mpc and
grid sizes of at least several thousand cells per side. These simulation 
parameters are now just becoming achievable on current hardware. These are of 
course only very rough estimates and they only relate to the 21-cm sky maps 
and statistical studies (e.g. power spectra, or PDFs). For other purposes, for 
example 21-cm absorption studies against bright radio galaxies (see e.g. 
Ciardi et al. (2015) reference: PoS(AASKA14)006), a much higher simulation 
resolution will be 
required, corresponding to frequency channels of $\sim$~kHz, or cells of 
$\sim10$~kpc comoving. This is unrealistic when combined with $\sim500$~Mpc 
volumes unless adaptive-mesh refinement (AMR) techniques are employed, or 
else smaller simulation volumes have to be used for such work.


Apart from the purely instrumental requirements discussed above, the 
simulation parameters are also dictated by the characteristic scales of the
reionization itself and the various processes that occur at different stages
and will be of interest to study. Before the first UV sources form, the 21-cm
fluctuations are dictated by the underlying density fluctuations, the 
temperature of the IGM, as well as additional effects like baryon-dark matter
displacement \cite[][see also Maio et al. 2015, reference: 
PoS(AASKA14)009]{2010PhRvD..82h3520T} and star formation suppression in 
minihaloes due to Lyman-Werner bands photons. Once the 
first ionizing sources appear, they propagate ionization fronts into the IGM 
initially forming small, local HII regions, which then quickly percolate and
merge locally into larger ones. The same first sources also produce copious 
amounts of soft-UV radiation and likely some X-rays, which heat the IGM gas
and decouple its spin temperature from the CMB, making it detectable. 
Outside dense regions/very high redshifts, the 21-cm line decoupling from the 
CMB also requires sufficiently strong Lyman-$\alpha$ background. 
All these different processes impose some characteristic scales on the 21-cm 
fluctuations. These are typically long-range modulations ranging from tens of 
Mpc (Lyman-$\alpha$), to $\sim100$~Mpc (baryon-dark matter displacement and 
Lyman-Werner) and up to hundreds of Mpc (X-rays). Later-on, likely below 
redshift $\sim10$ when the reionization becomes more widespread and is driven 
by larger, atomically-cooling halos, the 21-cm fluctuations due to these early 
backgrounds become less significant, baryon-dark matter displacement decreases 
and the dominant process determining the 21-cm fluctuations become the 
ionization patchiness. Its typical scales and geometry depend on the abundance 
and clustering of the main sources driving the process. Low-mass 
galaxies are exponentially more abundant, but are also liable to get their 
star formation suppressed by the Jeans mass filtering due to photoheathing,
as well as other radiative and mechanical feedback mechanisms. The details of 
these processes are still not firmly established and are currently subject of 
active research.

\begin{figure}
\includegraphics[width=.5\textwidth]{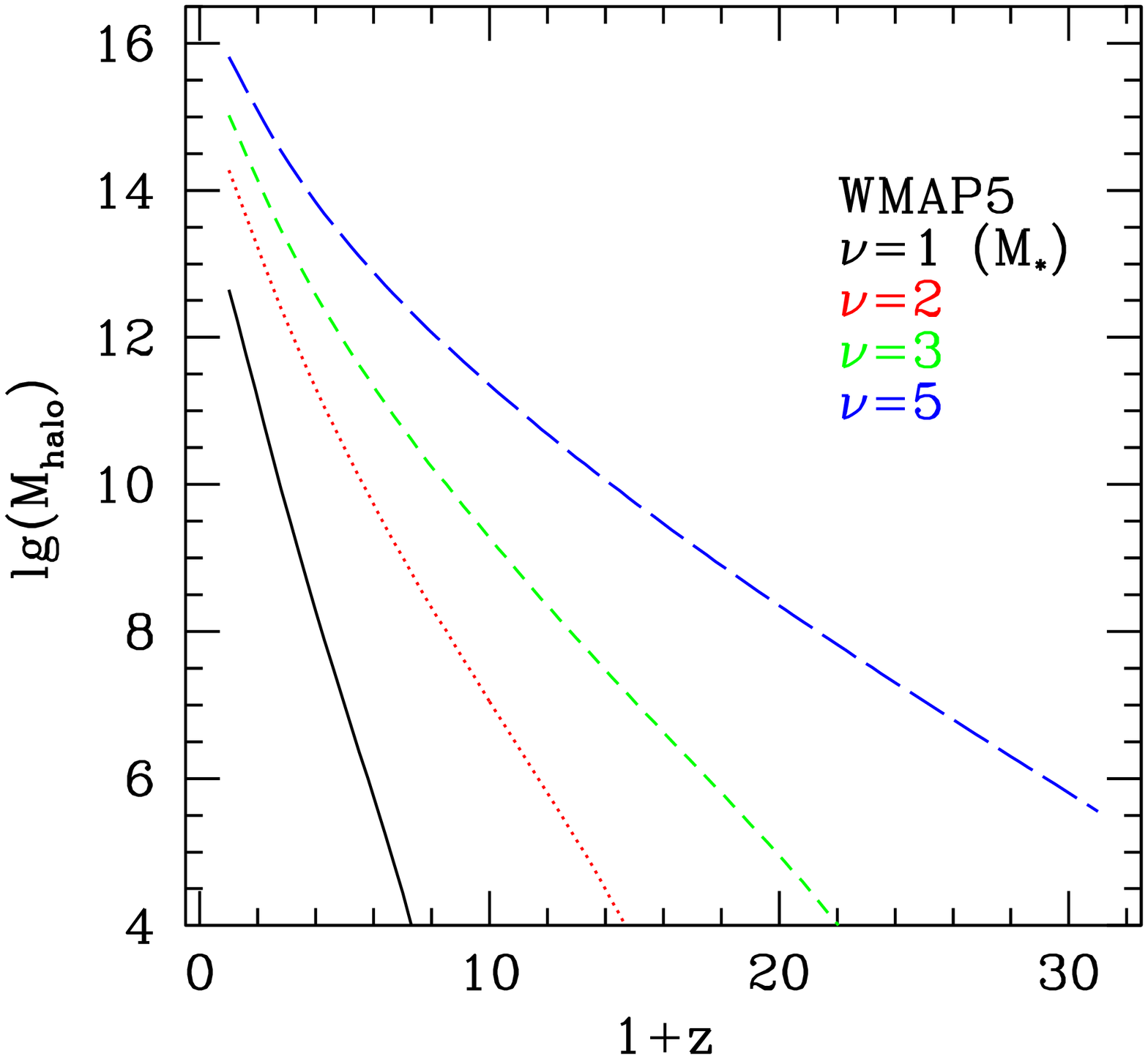}
\includegraphics[width=.5\textwidth]{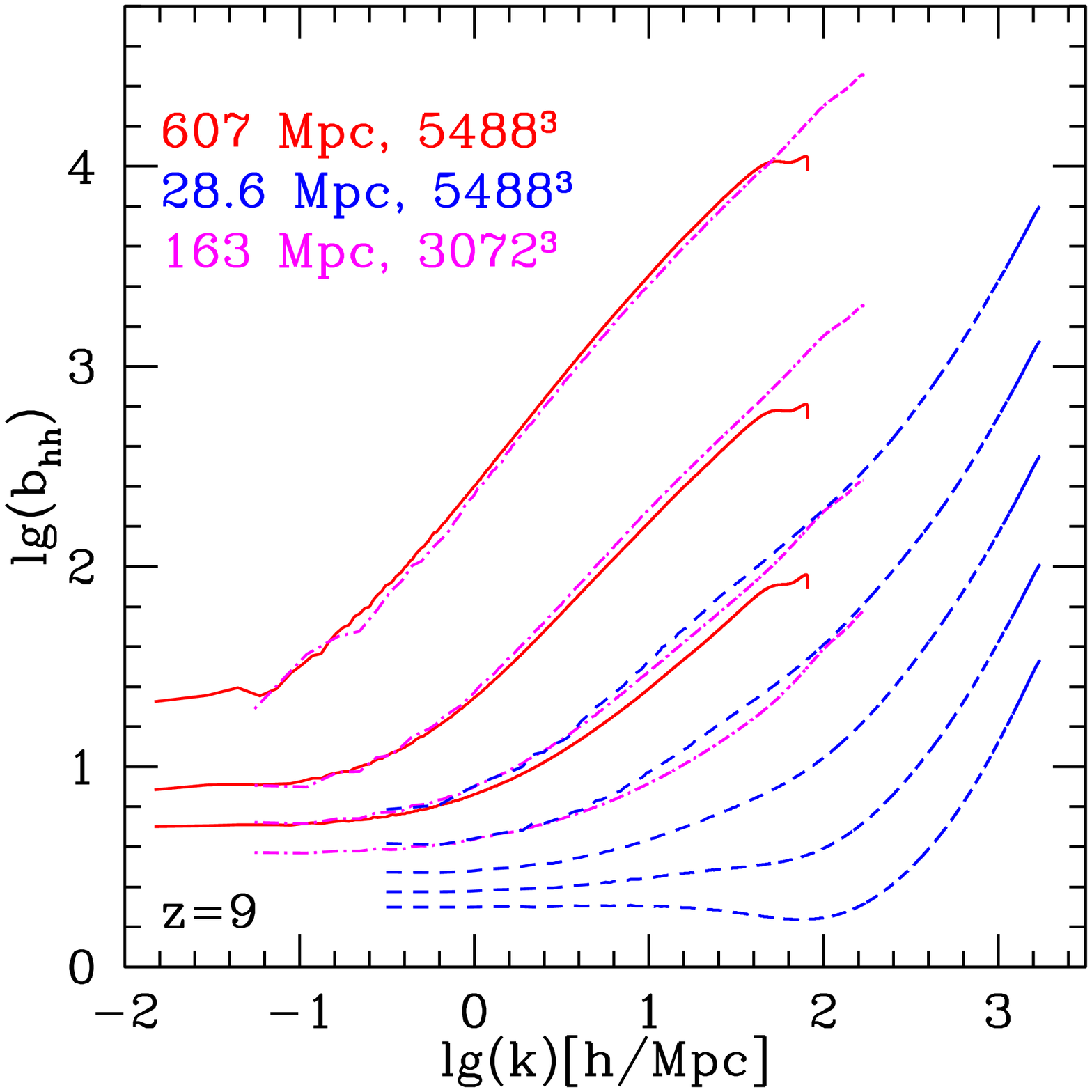}
\caption{(left) How common are halos of different mass at high redshift?
The lines show the mass vs redshift for $\nu-\sigma$ fluctuations in the
Gaussian statistics, where $\nu=\delta_c/\sigma(z=0,M)=1$ (i.e. M*, typical 
halo; black, solid), $\nu=2$ (red, dotted), $\nu=3$ (green, short-dashed) 
and $\nu=5$ (blue, long-dashed) \citep{iliev12}.
(right) The halo bias, $b_{\rm hh}^2=\Delta_{\rm hh}/\Delta_\rho$, at 
redshift $z=9$ for three cosmological volumes at different resolution, 
as labeled. Lines are for haloes binned by decades of mass (bottom to 
top curve) 
$10^5M_\odot\leq M_{\rm halo}<10^6M_\odot$, 
$10^6M_\odot\leq M_{\rm halo}<10^7M_\odot$, 
$10^7M_\odot\leq M_{\rm halo}<10^8M_\odot$, 
$10^8M_\odot\leq M_{\rm halo}<10^9M_\odot$, 
$10^9M_\odot\leq M_{\rm halo}<10^{10}M_\odot$, 
$10^{10}M_\odot\leq M_{\rm halo}<10^{11}M_\odot$,
$10^{11}M_\odot\leq M_{\rm halo}<10^{12}M_\odot$,
and $10^{12}M_\odot\leq M_{\rm halo}$.}
\label{halos:fig}
\end{figure}

\begin{figure}
\includegraphics[width=.5\textwidth]{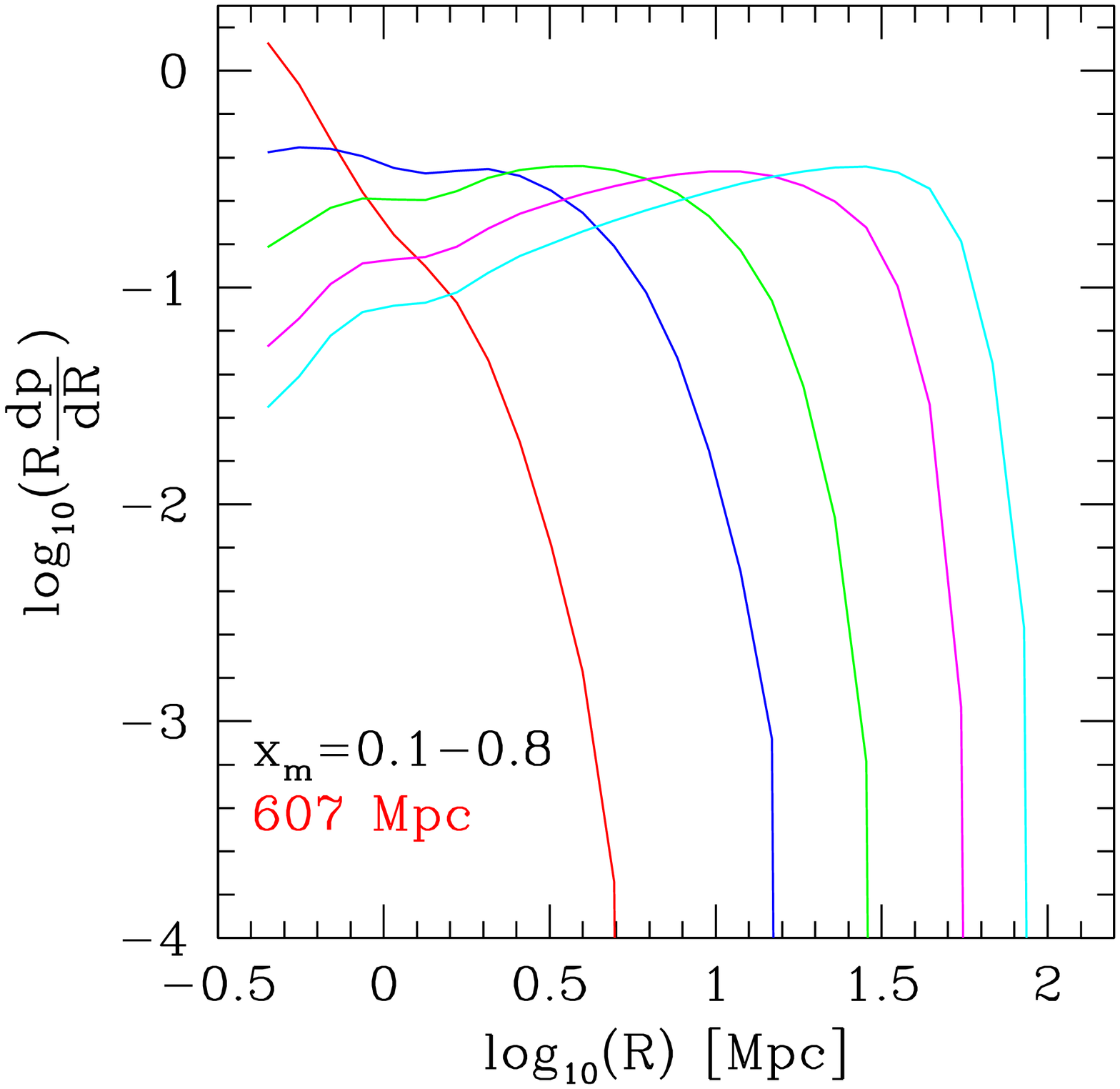}
\includegraphics[width=.5\textwidth]{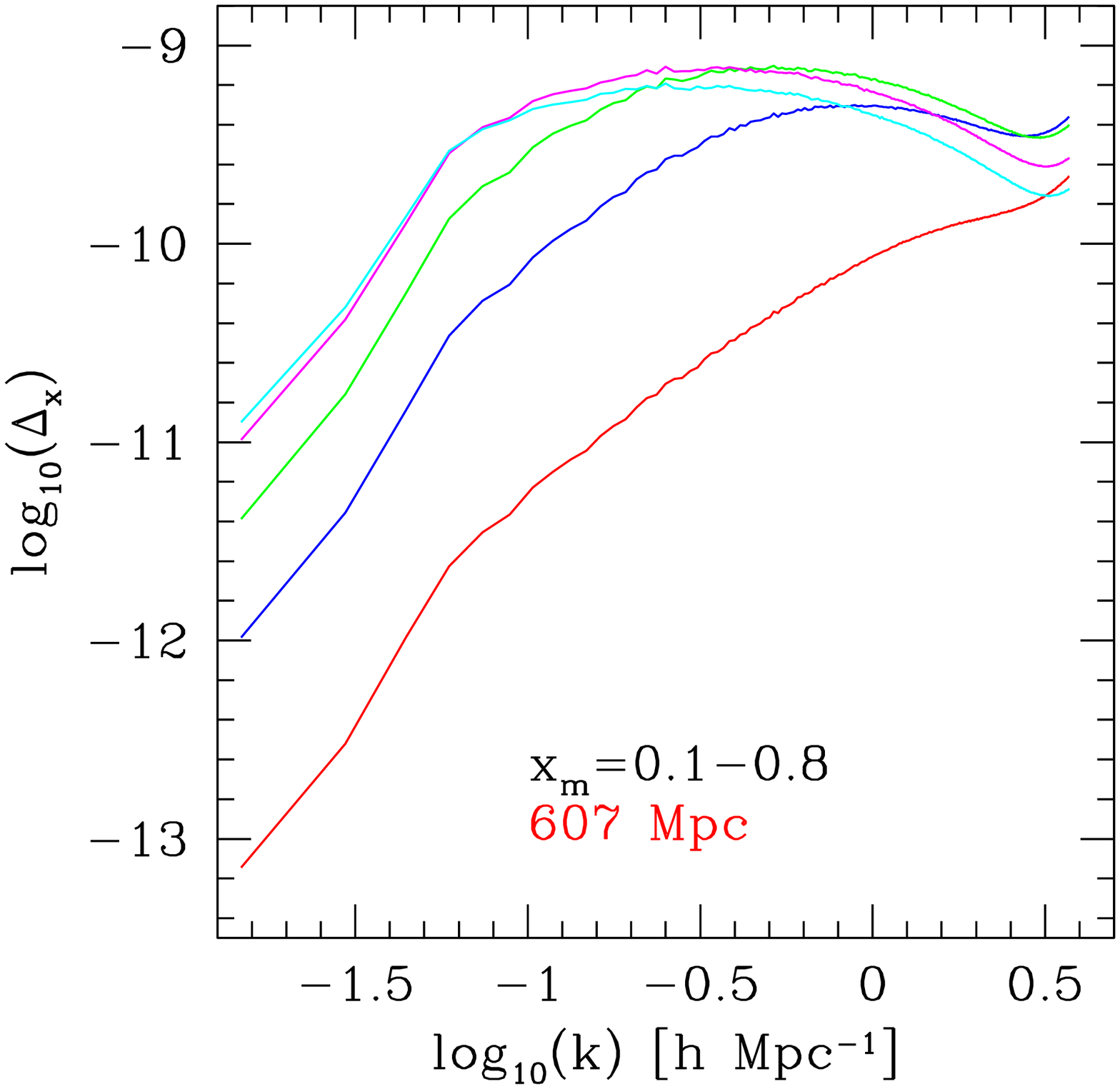}
\caption{(left panel) Probability distribution function per logarithmic radial 
bin, $R\frac{dp}{dR}$, spherical regions with radius $R$ as given by the 
Spherical Average method \citep{2007ApJ...654...12Z} based on the ionized 
distribution given by a reionization simulation with volume of 607 Mpc and 
$504^3$ grid \citep{2014MNRAS.439..725I}. Shown are results for different 
stages of the reionization process, for ionized fraction by mass $x_m=0.1$ 
to 0.8 (left to right). (right panel) Dimensionless power spectra, $\Delta_x$, 
of the volume-weighted 
ionized fraction, for H~II regions extracted from the same reionization 
simulation and at the same EoR stages (bottom to top).}
\label{sizes:fig}
\end{figure}

The ionized regions growth continues throughout reionization (see 
Fig.~\ref{evolution:fig} for an example). The characteristic patch sizes are 
dictated by the abundance, clustering and typical luminosities of the dominant
ionizing sources, as well as various effects which modify the growth like 
recombinations in the IGM gas, Lyman-limit systems and other photon sinks.   
The intrinsic source clustering results from the statistics of the initial
Gaussian random noise density fluctuations and depends strongly on redshift
and the typical mass of the halos hosting the dominant sources. In $\Lambda$CDM
the cosmological structures form hierarchically, with the smallest ones forming
first and then growing and merging over time to form larger ones. Consequently,
low-mass galaxies likely dominated the ionizing photon output throughout 
reionization. Figure~\ref{halos:fig} (left) shows the relative abundance of 
halos of different mass vs. redshift. At very high redshifts all halos are 
rare ($\nu\gg1$) and only below redshift $z\sim15$ the low-mass galaxies 
($M=10^8-10^{10}M_\odot$) become somewhat more common ($\nu=2-3$). Consequently,
these halos are strongly clustered, with bias with respect to the underlying 
mass distribution well above 1 (Figure~\ref{halos:fig}, right). This strong 
clustering yields quick percolation of the individually small HII regions 
around each relatively weak source and thus the rapid growth to much larger 
scales. Furthermore, there is a notable power in the density fluctuations at
fairly large scales, which required large simulation volumes to account for
properly \citep{mesinger07,2014MNRAS.439..725I}. This additional power means that local 
photon output is modulated significantly, which is not modelled correctly in
volumes smaller than about $\sim100\,\rm Mpc/h$ per side, resulting in 
artificial suppression of the redshifted 21-cm fluctuations, among others.
Different measures of the characteristic scales of the patchiness all yield
typical peak sizes of tens of comoving Mpc (see Figure~\ref{sizes:fig}), 
which corresponds to tens of arcmin \citep[see also][]{sobacchi14}. In summary,
accurate modelling of all these physical processes again points to requirements
of large, at least several hundred Mpc per side, volumes and sufficient resolution
to reliably identify, either directly or sub-grid, the main sources and sinks 
driving the reionization process.

%
%
%

\subsection{Computational Requirements}        

As mentioned above, reliable guidance and interpretation of the observations 
in terms of deriving the properties of the first galaxies requires the creation
of large libraries of models sampling the available parameters space.
At present a few full radiative transfer codes can be scaled up to the volumes 
and resolutions required for full SKA1 EoR simulation discussed in the previous 
section. This can only be achieved using massively-parallel approach (in some 
cases using accelerators like Graphical Processing Units, GPUs, and Intel Phi
multicore processors) on the largest available computers
and a single simulation of this size requires up to tens of millions of
core-hours. Even with the expected advances in computing technology it is 
unlikely that a large number of such simulations could be practically performed.
It is therefore most likely that more approximate methods like semi-numerical
modelling would have to be employed, guided by fewer detailed full simulations.

Both approaches also require large amounts of data storage. The large N-body 
structure formation simulations (needed for precision and accuracy even in 
semi-numerical models, as explained in \S~\ref{seminanalyt:sect}), alone 
require up to several PB of storage per simulation, with further significant 
storage required for the reionization data itself. Accessing and using this 
data within the community will require also investment in databases and other 
efficient methods for data sharing.   

\section{Summary}

We presented an overview of the current status and future prospects for 
simulations and modelling of the Epoch of Reionization, with focus on the 
specific requirements for SKA1-LOW. As discussed in detail, extensive modeling 
is particularly important for Cosmic Dawn and EoR science compared to other 
areas of study because as yet there is very limited direct observational data 
to guide us. Two basic approaches are available - full numerical simulations 
and semi-analytical/semi-numerical modeling. The simulations can handle the  
nonlinear dynamics, feedback effects and complex geometry, but are fairly 
computationally expensive. This drawback is being alleviated by innovative, 
efficient radiative transfer methods, some of which are able to efficiently 
utilise the latest Petascale computing facilities and specialist hardware 
(e.g. GPU accelerators). This now allows simulations which were impossible just 
a few years ago, with volumes and resolution matching both the intrinsic scales 
of the reionization process and the FOV and resolution expected for the SKA1 
EoR experiment. Nonetheless, full simulations remain quite expensive and 
consequently a significant effort has been invested in the development of 
semi-numerical modelling and other faster modelling methods. These approaches 
are by nature more approximate, generally use simplified physics and are 
missing the non-linear effects. However, when they are carefully constructed 
they still can retain many of the key features, while at the same time 
introducing the option of constructing large libraries of models to be used 
for interpreting the data, which is not feasible for full simulations. Much 
additional work is still required on these models, as well as on the numerical 
simulations, particularly in adding important physical processes which are 
still missing and in calibration and verification of semi-analytical models 
against numerical simulations. We have also outlined the requirements for 
SKA1-LOW-specific simulations and provided estimates on the resources 
which will be required for this effort.


\bibliographystyle{mn2e} 



\end{document}